\documentclass{aa}
\input epsf

\begin{document}

   \thesaurus{09(09.01.1;09.16.1;10.01.1)}
   \title{Abundances of Recently Discovered Planetary Nebulae Towards the Galactic 
Bulge\thanks{Based
on observations made at the European Southern Observatory (Chile) and Laborat\'orio
Nacional de Astrof\'\i sica (Brasil)}}

   \author{A.V. Escudero 
          \inst{1}
          \and
          R.D.D. Costa
          \inst{1}
          }

   \offprints{R.D.D. Costa}

   \institute{Departamento de Astronomia, IAG/USP, C.P. 3386, 01060-970 
S\~ao Paulo - SP, Brasil\\
email: escudero@iagusp.usp.br 
email: roberto@iagusp.usp.br 
}
   \date{Received 30 April 2001; Accepted 24 September 2001}
 
   \authorrunning{A.V. Escudero \& R.D.D. Costa}
   \titlerunning{ Abundances of PNe towards the Bulge}

   \maketitle

   \begin{abstract}
In this work we report spectrophotometric observations of a planetary nebula
sample towards the galactic bulge. A total of 45 PNe was observed and their 
physical parameters (electron density and temperature) were derived. Ionic
abundances were calculated using a three-level atom model and 
abundances were derived through ionisation correction factors. 
Results show low abundance objects at high galactic latitudes,
indicating a possible vertical gradient inside the bulge. A few objects
with low N/O ratio were found, which could originated from old, low mass
progenitors.

   \end{abstract}

%

\section{Introduction}

The galactic bulge is a structure whose properties have been extensively
studied in recent years. In particular, abundance studies show many
aspects of bulge populations, sometimes contradictory: while earlier 
works (e.g. Rich 1988) 
indicate abundances up to [Fe/H] $\sim$ 0.30, other more recent studies 
(McWilliam \& Rich 1994, Ibata \& Gilmore 1995) indicate subsolar values, 
derived from K giants. Using spectra of integrated light from the galactic
bulge in Baade's window, Idiart et al. (1996) derive a mean metallicity
[Fe/H] = $-$0.02, which is consistent with the mean values found by Rich (1988)
and McWilliam
\& Rich (1994). Comprehensive revisions of abundance data and the current 
ideas on Galaxy 
evolution can be found in van den Bergh (1996) and McWilliam (1997). 

In this context, abundances from planetary nebulae (PNe) play an 
important role. The derivation of nebular abundances of light elements 
like O, S, Ar from the flux of forbidden emission lines is quite convenient 
at the large distances and strong reddening of the galactic bulge. Abundances 
from these objects reflect the state of the interstellar medium at the 
progenitor formation epoch,
and therefore can be related directly to stellar abundances. On the other hand, 
He and N
abundances are modified along the stellar evolution of the progenitor 
and can be related to their masses (and age).

Some previous works (Ratag et al. 1992, 1997; Cuisinier et al. 2000)
give abundances of different samples of bulge PNe. In both cases, objects were 
selected from the Acker et al. (1992) catalogue, and results indicate an 
abundance pattern comparable to their disk counterparts. A point yet to be tackled
is the possibility of the existence of a chemical abundance gradient within
the bulge. There are several suggestions for the existence of a 
chemical abundance gradient within the bulge, as proposed by Minitti et al. (1995),
Frogel et al. (1999) or Tiede et al. (1995). In this last work they found
evidence of a gradient in Fe/H for the region $-$12 $\le b \le$ $-$3.

As part of a long-term program of derivation of nebular abundances,
we have carried out observations of southern planetary nebulae (see, e.g.
Costa et al. 1996, 2000  and references therein)
aiming to derive their nebular physical parameters and chemical abundances.
In this work, we report spectroscopic observations for a sample of 44 PNe 
towards the galactic bulge. From this sample, chemical abundances were
derived. Objects were selected from new samples recently 
published
(Kohoutek 1994; Beaulieu et al. 1999), and with these results we intend
to enlarge the sample of bulge PNe with known chemical abundances, to 
obtain the abundance 
distribution of the intermediate mass star population in the galactic bulge.
These results give important clues for the chemical enrichment and
evolution of the galactic bulge.

In section 2 we present our observation and reduction procedures, in 
sect. 3
physical parameters and chemical abundances derivation are derived,
in sect. 4 results are discussed and general
conclusions are drawn.


\section{Observations and data reduction}

\subsection{Observations}

Observation were made at the 1.60m telescope of the National Laboratory for 
Astrophysics (LNA, Bras\'opolis, Brazil) and at the 1.52m of the European Southern 
Observatory (ESO/La Silla, Chile). In both telescopes, Cassegrain Boller \&
Chivens spectrographs were used, with reciprocal dispersions of 4.4 \AA/pixel at 
LNA and 2.2 \AA/pixel at ESO. A long east-west slit was used in all observations.
In both observatories, slits of 2 arcsecs width were used.
The log of the observations and identification of the objects
are given in table 1.
Each object was observed at least two times and line fluxes were
derived from the average of all measures for the same object. 

Data reduction was performed using the IRAF package,
following the standard procedure for longslit spectra: correction of
bias, flat-field, extraction, wavelength and flux calibration. For
flux calibration, we observed at least three spectrophotometric standard stars 
each night. Atmospheric extinction was corrected through mean coefficients 
derived for each observatory.

Line fluxes were calculated adopting gaussian profiles, and a
gaussian de-blending routine was used when necessary. 
Line fluxes are given in tables 2 and 3 using the usual scale of F(H$\beta$)=100.
The H$\beta$ fluxes through the slit, in physical units (erg/cm$^2$.s), and corrected 
for interstellar and atmospheric reddening are listed in table 4.
Interstellar reddening was derived from the
Balmer ratio H$\alpha$/H$\beta$, assuming Case B (Osterbrock 1989), 
using the interstellar extinction law by Cardelli et al. (1989)  and
adopting R$_V$=3.1 for all objects. 
E(B-V) values derived for the nebulae are given in table 4.

\begin{table}
\caption[]{Log of the observations}
\begin{flushleft}
\begin{tabular}{lrrrl}
\hline\noalign{\smallskip}
PNe & PN G / PK & Num. & Date of Obs. & Place\\
\noalign{\smallskip}
\hline\noalign{\smallskip}
K5-1 & PK 000+04.3 & 2 & 20-Jul-99 & LNA\\
K5-3 & PN G 002.6+05.5 & 2 & 31-May-98 & LNA\\
 & & 2 & 16-Aug-99 & ESO\\
K5-4 & PN G 351.9-01.9 & 2 & 31-May-98 & LNA\\
 & & 2 & 15-Aug-99 & ESO\\
K5-5 & PN G 001.5+03.6 & 2 & 20-Jul-99 & LNA\\
K5-6 & PN G 003.6+04.9 & 2 & 20-Jul-99 & LNA\\
K5-7 & PN G 003.1+04.1 & 3 & 19-Jul-99 & LNA\\
 & & 2 & 16-Aug-99 & ESO\\
K5-9 & PK 355-01.1 & 2 & 19-Jul-99 & LNA\\
K5-11 & PK 002+02.1 & 2 & 31-May-98 & LNA\\
 & & 2 & 15-Aug-99 & ESO\\
K5-12 & PK 353-03.1 & 2 & 19-Jul-99 & LNA\\
K5-13 & PK 002+02.2 & 2 & 31-May-98 & LNA\\
 & & 2 & 16-Aug-99 & ESO\\
K5-14 & PN G 003.9+2.6 & 1 & 18-Jul-99 & LNA\\
 & & 1 & 19-Jul-99 & LNA\\
K5-16 & PN G 002.8+01.8 & 1 & 31-May-98 & LNA\\
 & & 2 & 18-Jul-99 & LNA\\
K5-17 & PN G 004+02.2 & 2 & 18-Jul-99 & LNA\\
K5-19 & PN G 005.1+02.0 & 1 & 18-Jul-99 & LNA\\
K5-20 & PN G 356.8-03.0 & 3 & 18-Jul-99 & LNA\\
 & & 2 & 16-Aug-99 & ESO\\
SB 01 & PN G 000.1-08.0 & 2 & 24-Jun-00 & LNA\\
SB 02 & PN G 000.5-05.3 & 2 & 21-Jun-00 & LNA\\
SB 03 & PN G 000.7-06.1 & 2 & 23-Jun-00 & LNA\\
SB 04 & PN G 001.1-06.4 & 2 & 24-Jun-00 & LNA\\
SB 06 & PN G 001.6-05.9 & 2 & 23-Jun-00 & LNA\\
SB 12 & PN G 005.4-06.1 & 2 & 24-Jun-00 & LNA\\
SB 15 & PN G 009.3-06.5 & 2 & 23-Jun-00 & LNA\\
SB 17 & PN G 011.1-07.9 & 2 & 17-Aug-99 & ESO\\
SB 18 & PN G 011.4-07.3 & 1 & 23-Jun-00 & LNA\\
 & & 2 & 25-Jun-00 & LNA\\
SB 19 & PN G 014.4-06.1 & 2 & 17-Aug-99 & ESO\\
SB 20 & PN G 014.8-08.4 & 2 & 22-Jun-00 & LNA\\
SB 21 & PN G 016.0-07.6 & 2 & 16-Aug-99 & ESO\\
SB 24 & PN G 017.5-09.2 & 2 & 21-Jun-00 & LNA\\
SB 25 & PN G 341.0+09.4 & 2 & 19-Aug-99 & ESO\\
SB 26 & PN G 341.7-06.0 & 2 & 21-jun-00 & LNA\\
SB 28 & PN G 342.3-06.0 & 2 & 18-Aug-99 & ESO\\
SB 30 & PN G 343.9-05.8 & 2 & 23-Jun-00 & LNA\\
SB 31 & PNG 347.9-06.0 & 2 & 18-Aug-99 & ESO\\
SB 32 & PNG 349.7-09.1 & 2 & 20-Aug-99 & ESO\\
 & & 1 & 21-Jun-00 & LNA\\
SB 33 & PNG 351.2-06.3 & 2 & 24-Jun-00 & LNA\\
SB 34 & PNG 351.5-06.5 & 1 & 18-Aug-99 & ESO\\
 & & 1 & 19-Aug-99 & ESO\\
SB 35 & PNG 351.7-06.6 & 2 & 20-Aug-99 & ESO\\
 & & 2 & 24-Jun-00 & LNA\\
SB 37 & PNG 352.6-04.9 & 2 & 19-Aug-99 & ESO\\
SB 38 & PNG 352.7-08.4 & 2 & 22-Jun-00 & LNA\\
SB 42 & PNG 355.3-07.5 & 2 & 22-Jun-00 & LNA\\
SB 44 & PNG 356.0-07.4A & 2 & 18-Aug-99 & ESO\\
SB 50 & PNG 357.3-06.5 & 2 & 22-Jun-00 & LNA\\
SB 52 & PNG 358.3-07.3 & 2 & 22-Jun-00 & LNA\\
SB 53 & PNG 358.7-05.1 & 2 & 23-Jun-00 & LNA\\
SB 55 & PNG 359.4-08.5 & 2 & 17-Aug-99 & ESO\\
\noalign{\smallskip}
\hline
\end{tabular}
\end{flushleft}
\end{table}

All objects from Beaulieu et al. (1999) and Kohoutek's (1994) table 5
were observed. Those not present in table 1 do not have a S/N satisfactory
for plasma diagnostics. In particular, objects K5-8, SB-05, SB-27 and SB-29
do not seem to be PNe.

\begin{table*}
\caption[]{Line fluxes}
\begin{flushleft}
\begin{tabular}{lrrrrrrrrrrrr}
\hline\noalign{\smallskip}
Line & K5-1 & K5-3 & K5-4 & K5-5 & K5-6 & K5-7 & K5-9 & K5-11 & K5-12 & K5-13 & K5-14 & K5-16\\
\noalign{\smallskip}
\hline\noalign{\smallskip}
$[OII]\lambda$3726+29 & - & 63.0 & 133.7 & - & - & - & - & - & - & 117.3 & - & -\\
$[NeIII]\lambda$3869 & - & 126.1 & 172.6 & - & - & 118.7 & - & - & 51.9 & 196.3 & 38.1 & -\\
$[NeIII]\lambda$3967 & - & 55.3 & 62.1 & - & - & - & - & - & 23.4 & 75.8 & - & -\\
$H\gamma\lambda$4340 & 65.5 & 56.2 & 57.3 & 91.6 & 65.5 & 74.6 & - & - & 75.5 & 54.2 & 41.7 & - \\
$[OIII]\lambda$4363 & 7.9 & 12.9 & 6.9 & 30.6 & 25.8 & 26.8 & - & 2.6 & 9.8 & 20.3 & 37.5 & -  \\
$HeI\lambda$4472 & 11.9 & 10.1 & 6.2 & 18.6 & 10.2 & - & 8.7 & - & - & 11.8 & 4.2 & -  \\
$HeII\lambda$4686 & - & 51.5 & - & - & 102.2 & 106.6 & 18.3 & 21.4 & 71.9 & 73.8 & 26.2 & 15.8  \\
$H\beta\lambda$4861 & 100.0 & 100.0 & 100.0 & 100.0 & 100.0 & 100.0 & 100.0 & 100.0 & 100.0 & 100.0 & 100.0 & 100.0  \\
$[OIII]\lambda$4959 & 168.7 & 276.7 & 357.9 & 283.0 & 310.6 & 339.6 & 225.4 & 172.1 & 402.8 & 441.6 & 571.2 & 287.1  \\
$[OIII]\lambda$5007 & 507.7 & 818.9 & 1061.5 & 843.3 & 915.5 & 1024.7 & 659.3 & 496.4 & 1201.3 & 1328.9 & 1663.8 & 859.1  \\
$HeII\lambda$5412 & - & 4.6 & 0.4 & - & 13.6 & 6.0 & 1.9 & - & 5.4 & 6.6 & 2.4 & 2.3  \\
$[NII]\lambda$5755 & 0.9 & 0.5 & 1.9 & 2.0 & 2.9 & - & 0.7 & 3.1 & 0.7 & 4.3 & 6.4 & 6.1  \\
$HeI\lambda$5876 & 17.5 & 11.6 & 17.4 & 9.7 & 3.6 & 8.6 & 21.1 & 25.1 & 7.4 & 11.8 & 15.5 & 18.3  \\
$[OI]\lambda$6300 & 0.6 & 1.0 & 6.2 & 2.9 & 1.7 & 12.9 & 2.3 & 8.3 & 2.4 & 6.4 & 14.0 & 31.1  \\
$[SIII]\lambda$6312 & 0.7 & 1.4 & 1.8 & 1.3 & 1.8 & - & 0.6 & 1.1 & 1.9 & 6.0 & 3.9 & 5.0  \\
$[NII]\lambda$6548 & 4.2 & 10.7 & 25.4 & 3.7 & 2.3 & 9.5 & 11.7 & 107.9 & 6.8 & 111.5 & 58.1 & 164.8  \\
$H\alpha\lambda$6563 & 285.2 & 285.2 & 285.2 & 285.3 & 285.2 & 285.2 & 285.3 & 285.2 & 285.2 & 285.2 & 285.2 & 285.3  \\
$[NII]\lambda$6584 & 18.1 & 23.8 & 71.8 & 20.5 & 7.4 & 19.1 & 41.8 & 333.5 & 25.1 & 345.7 & 208.6 & 501.8  \\
$HeI\lambda$6678 & 5.1 & 3.7 & 4.4 & 4.2 & 2.4 & 4.6 & 6.4 & 7.6 & 2.4 & 5.2 & 4.4 & 5.7  \\
$[SII]\lambda$6716 & 1.4 & 3.4 & 3.2 & 1.5 & 2.5 & 5.4 & 2.7 & 25.4 & 2.8 & 28.9 & 8.6 & 59.5  \\
$[SII]\lambda$6731 & 1.9 & 4.4 & 6.1 & 3.0 & 2.9 & 4.6 & 4.4 & 26.8 & 5.0 & 35.9 & 16.6 & 62.3  \\
$[ArIII]\lambda$7135 & 15.5 & 11.0 & 15.3 & 10.4 & 8.9 & 24.7 & 16.1 & 19.5 & 91.4 & 40.4 & 32.3 & 24.5  \\
$[OII]\lambda$7320+30 & 2.4 & 2.8 & 11.7 & 10.0 & 2.0 & 3.7 & 3.3 & 6.0 & 3.3 & 5.2 & 26.5 & 10.8  \\
\noalign{\smallskip}
\hline
\noalign{\smallskip}
Line & K5-17 & K5-19 & K5-20 & SB01 & SB02 & SB03 & SB04 & SB06 & SB12 & SB15 & SB17 & SB18\\
\noalign{\smallskip}
\hline\noalign{\smallskip}
$[OII]\lambda$3726+29  & - & - & 33.1 & 109.0 & 81.7 & 41.2 & - & 458.8 & - & 56.8 & 553.6 & 154.3 \\
$[NeIII]\lambda$3869  & - & - & 87.3 & 152.2 & 25.9 & 89.1 & 78.0 & 82.2 & - & 189.6 & - & 70.7  \\
$[NeIII]\lambda$3967  & - & - & 48.4 & 81.3 & 50.8 & - & - & 75.4 & - & - & 19.5 & 35.7  \\
$H\gamma\lambda$4340  & 32.4 & - & 47.3 & 35.8 & 53.4 & 49.1 & 49.3 & 36.5 & 37.3 & 46.5 & 53.8 & 35.5  \\
$[OIII]\lambda$4363  & 23.7 & 54.4 & 7.0 & 7.6 & 8.6 & 36.5 & 15.4 & 19.0 & 35.9 & 46.7 & - & 27.6  \\
$HeI\lambda$4472  & - & - & 7.9 & 22.4 & 18.1 & 3.8 & - & 12.5 & - & - & - & 12.7  \\
$HeII\lambda$4686  & - & 95.8 & - & 37.7 & 32.2 & 59.5 & 21.6 & 50.4 & 72.1 & 51.3 & - & 64.3  \\
$H\beta\lambda$4861  & 100.0 & 100.0 & 100.0 & 100.0 & 100.0 & 100.0 & 100.0 & 100.0 & 100.0 & 100.0 & 100.0 & 100.0  \\
$[OIII]\lambda$4959  & 408.2 & 422.1 & 269.7 & 164.9 & 159.9 & 208.6 & 181.0 & 154.8 & 322.6 & 442.5 & - & 339.2  \\
$[OIII]\lambda$5007  & 1247.1 & 1173.4 & 808.3 & 540.1 & 477.8 & 616.5 & 522.1 & 504.5 & 799.8 & 1300.1 & 13.4 & 972.8  \\
$HeII\lambda$5412  & 0.9 & - & - & - & - & 3.7 & - & 6.6 & - & 27.0 & - & -  \\
$[NII]\lambda$5755  & - & 1.6 & 1.8 & - & 2.6 & 1.6 & - & - & - & - & 6.5 & -  \\
$HeI\lambda$5876  & 14.1 & 10.4 & 17.6 & 6.7 & 18.3 & 14.2 & 11.7 & 10.1 & 11.9 & 11.8 & 18.6 & 6.0  \\
$[OI]\lambda$6300  & 2.4 & - & 3.1 & 7.0 & 13.6 & - & - & - & - & - & 37.4 & 3.7  \\
$[SIII]\lambda$6312  & 1.2 & - & 0.5 & - & - & - & - & - & - & - & - & -  \\
$[NII]\lambda$6548  & 5.7 & 13.5 & 2.0 & 6.7 & 28.4 & 12.1 & 42.7 & 176.5 & - & - & 94.5 & -  \\
$H\alpha\lambda$6563  & 285.2 & 285.2 & 285.2 & 285.1 & 285.1 & 285.1 & 285.0 & 285.1 & 285.0 & 285.1 & 285.1 & 285.0  \\
$[NII]\lambda$6584  & 20.8 & 42.8 & 4.5 & 17.5 & 78.5 & 42.4 & 116.4 & 531.6 & 7.5 & 1.6 & 256.6 & 3.6  \\
$HeI\lambda$6678  & 3.7 & 2.5 & 4.6 & 5.0 & 4.6 & 5.6 & 4.8 & 6.6 & 17.2 & 8.8 & 6.0 & 2.8  \\
$[SII]\lambda$6716  & 2.0 & 7.2 & 0.8 & 5.0 & 12.7 & 7.0 & 15.1 & 14.4 & - & - & 57.5 & 1.8  \\
$[SII]\lambda$6731  & 4.0 & 8.5 & 0.8 & 6.6 & 10.3 & 6.2 & 11.7 & 10.9 & - & 6.0 & 44.1 & 1.5  \\
$[ArIII]\lambda$7135  & 11.1 & 14.9 & 8.2 & 4.3 & 10.9 & 10.5 & 8.2 & 20.8 & 15.2 & 2.2 & 1.2 & 2.9  \\
$[OII]\lambda$7320+30  & 4.8 & 3.2 & 2.9 & 4.0 & 5.8 & - & 5.2 & 18.1 & - & 8.9 & 13.0 & 5.6  \\
\noalign{\smallskip}
\hline
\end{tabular}
\end{flushleft}
\end{table*}

\begin{table*}
\caption[]{Line fluxes (continued)}
\begin{flushleft}
\begin{tabular}{lrrrrrrrrrrrr}
\hline\noalign{\smallskip}
Line & SB19 & SB20 & SB21 & SB24 & SB25 & SB26 & SB28 & SB30 & SB31 & SB32 & SB33 & SB34\\
\noalign{\smallskip}
\hline\noalign{\smallskip}
$[OII]\lambda$3726+29  & 42.4 & 59.8 & - & 39.6 & 209.1 & 29.2 & - & 53.1 & 14.4 & 398.6 & 103.7 & -  \\
$[NeIII]\lambda$3869  & 63.1 & 254.1 & 51.9 & - & 97.1 & 126.3 & 55.9 & 74.8 & 84.5 & 60.2 & 24.5 & 79.6  \\
$[NeIII]\lambda$3967  & - & - & 30.6 & 35.8 & - & 53.9 & 25.5 & 32.3 & 36.2 & 34.2 & - & -  \\
$H\gamma\lambda$4340  & 31.9 & 36.4 & 54.6 & 21.6 & 49.9 & 40.1 & 66.2 & 52.1 & 63.6 & 47.0 & 51.1 & 68.5  \\
$[OIII]\lambda$4363  & 2.9 & 70.5 & 14.3 & 6.9 & 7.0 & 21.0 & 29.3 & 16.1 & 14.3 & 15.9 & 22.9 & 22.1  \\
$HeI\lambda$4472  & 7.6 & - & - & - & - & - & - & 3.6 & - & 9.4 & 28.9 & -  \\
$HeII\lambda$4686  & - & 30.2 & 116.9 & 90.7 & - & 94.5 & 116.2 & 46.6 & 117.0 & 25.7 & 17.5 & 140.4  \\
$H\beta\lambda$4861  & 100.0 & 100.0 & 100.0 & 100.0 & 100.0 & 100.0 & 100.0 & 100.0 & 100.0 & 100.0 & 100.0 & 100.0  \\
$[OIII]\lambda$4959  & 224.7 & 497.0 & 188.6 & 117.5 & 170.5 & 308.7 & 174.5 & 320.5 & 227.2 & 189.6 & 222.7 & 162.9  \\
$[OIII]\lambda$5007  & 648.7 & 1441.1 & 571.9 & 384.9 & 515.8 & 915.6 & 525.0 & 941.1 & 662.4 & 565.2 & 585.0 & 509.8  \\
$HeII\lambda$5412  & - & - & 7.9 & 3.2 & - & 7.4 & 6.7 & 3.6 & 7.3 & - & - & 21.5  \\
$[NII]\lambda$5755  & - & - & - & - & - & - & - & - & - & 5.0 & 11.9 & -  \\
$HeI\lambda$5876  & 22.5 & 13.7 & 4.0 & 5.8 & 18.1 & 3.0 & 1.1 & 10.5 & 1.3 & 15.1 & 10.7 & 2.4  \\
$[OI]\lambda$6300  & - & - & - & - & 9.3 & 2.4 & - & 4.2 & 0.9 & 32.1 & 31.8 & 7.9  \\
$[SIII]\lambda$6312  & 1.5 & - & - & - & 1.0 & 0.4 & - & 0.6 & 1.8 & 8.6 & - & -  \\
$[NII]\lambda$6548  & - & - & - & - & 107.0 & 2.5 & - & 6.4 & - & 37.6 & 77.0 & 7.9  \\
$H\alpha\lambda$6563  & 285.1 & 285.0 & 285.1 & 285.1 & 285.1 & 285.1 & 285.1 & 285.1 & 285.1 & 285.0 & 285.0 & 285.1  \\
$[NII]\lambda$6584  & 8.2 & 1.8 & 1.6 & 6.3 & 313.6 & 1.7 & 1.3 & 21.6 & 4.8 & 110.6 & 205.2 & 6.4  \\
$HeI\lambda$6678  & 8.5 & 2.8 & 1.4 & 6.8 & 4.2 & 2.5 & - & 2.8 & 1.8 & 7.9 & 10.2 & 1.3  \\
$[SII]\lambda$6716  & 2.7 & 3.9 & - & 2.1 & 38.4 & - & - & 1.2 & 1.0 & 32.8 & 32.3 & -  \\
$[SII]\lambda$6731  & 3.7 & 2.6 & - & 4.0 & 31.9 & - & - & 2.2 & 1.2 & 26.0 & 28.9 & -  \\
$[ArIII]\lambda$7135  & 19.8 & 3.0 & 7.8 & 4.8 & 19.6 & 5.4 & 4.4 & 5.3 & 5.5 & 8.8 & 15.7 & 4.6  \\
$[OII]\lambda$7320+30  & 14.6 & - & 3.0 & 10.3 & - & 1.0 & - & 6.0 & 1.2 & 15.9 & 12.9 & -  \\
\noalign{\smallskip}
\hline
\noalign{\smallskip}
Line &  SB35 & SB37 & SB38 & SB42 & SB44 & SB50 & SB52 & SB53 & SB55\\
\hline\noalign{\smallskip}
$[OII]\lambda$3726+29  & 29.8 & 38.4 & 6.0 & 35.8 & - & 72.5 & 33.4 & 43.0 & 15.4  \\
$[NeIII]\lambda$3869  & 73.0 & 132.2 & 63.1 & 29.1 & 25.1 & 62.9 & 68.2 & 71.1 & 56.6  \\
$[NeIII]\lambda$3967  & 18.9 & 47.3 & 13.4 & 26.9 & - & - & 37.5 & 39.0 & 24.0  \\
$H\gamma\lambda$4340  & 42.6 & 76.2 & 44.6 & 32.8 & 32.4 & 51.3 & 56.1 & 53.2 & 46.1  \\
$[OIII]\lambda$4363  & 8.3 & 32.5 & 26.1 & 3.0 & 6.7 & 20.0 & 12.1 & 23.7 & 17.0  \\
$HeI\lambda$4472  & 11.9 & - & - & 7.6 & - & - & 2.8 & - & -  \\
$HeII\lambda$4686  & - & 166.3 & 65.6 & 4.9 & 129.8 & 96.5 & 121.4 & 128.1 & 126.7  \\
$H\beta\lambda$4861  & 100.0 & 100.0 & 100.0 & 100.0 & 100.0 & 100.0 & 100.0 & 100.0 & 100.0  \\
$[OIII]\lambda$4959  & 266.2 & 401.8 & 303.1 & 89.0 & 188.0 & 228.5 & 228.6 & 312.7 & 123.7  \\
$[OIII]\lambda$5007  & 766.0 & 1213.7 & 888.2 & 245.0 & 627.5 & 685.7 & 674.9 & 904.5 & 353.0  \\
$HeII\lambda$5412  & - & 12.0 & 3.5 & - & 7.6 & 5.9 & 9.8 & 8.7 & 9.0  \\
$[NII]\lambda$5755  & 1.1 & - & - & 0.3 & - & - & 0.2 & 1.2 & -  \\
$HeI\lambda$5876  & 16.9 & - & 15.6 & 6.5 & 4.0 & 7.4 & 4.8 & 2.5 & 7.2  \\
$[OI]\lambda$6300  & - & 0.7 & - & - & - & - & 1.3 & 3.3 & -  \\
$[SIII]\lambda$6312  & - & 3.4 & - & - & - & - & 1.2 & 4.2 & -  \\
$[NII]\lambda$6548  & 3.4 & - & - & - & 15.1 & - & 0.8 & 5.6 & -  \\
$H\alpha\lambda$6563  & 285.1 & 285.1 & 285.0 & 285.1 & 285.0 & 285.0 & 285.0 & 285.1 & 285.1  \\
$[NII]\lambda$6584  & 2.7 & 13.9 & 2.2 & 10.8 & 19.8 & 14.0 & 5.6 & 16.1 & 3.2  \\
$HeI\lambda$6678  & 5.5 & 3.3 & 3.3 & 4.6 & 2.6 & 5.1 & 2.8 & 2.6 & 3.8  \\
$[SII]\lambda$6716  & 1.5 & 3.5 & 3.1 & 2.0 & 3.8 & 6.5 & 2.2 & 4.1 & 0.3 \\
$[SII]\lambda$6731  & - & 3.2 & - & 1.5 & 4.3 & 5.8 & 2.6 & 4.2 & 0.4  \\
$[ArIII]\lambda$7135  & 7.8 & - & 6.1 & 3.4 & 14.3 & 12.1 & 12.2 & 19.1 & 5.6  \\
$[OII]\lambda$7320+30  & 2.2 & 0.8 & 3.5 & 2.8 & 7.9 & - & 2.9 & - & 5.4  \\
\noalign{\smallskip}
\hline
\end{tabular}
\end{flushleft}
\end{table*}

\subsection {Physical parameters}

Electron densities were derived from the [SII] $\lambda$6716/$\lambda$6731 
ratio, while electron temperatures were derived from both 
[OIII] $\lambda$4363/$\lambda$5007 and [NII] $\lambda$5755/$\lambda$6584  
line ratios. When both temperatures were available, the [OIII] temperature
was used to derive ionic abundances  of species with higher ionisation
potentials as $\rm O^{+2},Ar^{+2},Ar^{+3},Ne^{+2}$, and the [NII] temperature
was used for lower potential lines such as $\rm O^+, N^+, S^+$.
When only one temperature was available, it was used for all ions.
Concerning densities, when [SII] lines could not be used for diagnostics,
an average value of 5000 cm$^{-3}$ was adopted. Derived values
are shown in table 4. Electron densities are in units of 10$^3$ cm$^{-3}$
and temperatures in units of 10$^4$ K. 

One of the main problems of observations towards the bulge is the high 
extinction, which affects mainly lower intensity lines. 
Lines [SII]$\lambda$6716+31 for example, are sometimes weak, but needed 
to derive electron densities. Other weak lines like [SIII]$\lambda$6312
or HeII$\lambda$5412 are also needed to derive the appropriate ionic abundances.
A more serious problem arises
from [OIII]$\lambda$4363 and [NII]$\lambda$5755 because abundances are strongly 
sensitive to electron temperature. Errors in these line fluxes will
contribute to the derived errors in the final abundances.

Examining the line fluxes for objects measured in both observatories, no
systematic differences were found, and we plot all the data together . 

As we have independent measures for each object, it was possible to
estimate the errors in line intensities. For this, we plot in fig. 1 all
line fluxes for all objects using reddened values, and estimate the error
for each line from its average and dispersion. From this distribution,
we derived the following expression for the correlation between errors and fluxes:

$$ log(\Delta F)=-5.67\pm 0.55 + (0.67\pm 0.04)log(F) $$

Where line fluxes are represented by F and their errors by $\Delta$F (in physical units). 
The full line in fig. 1 represents $\Delta$F = F. 

In order to estimate the errors in abundances, each line flux
had its value varied randomly 500 times within its respective error interval,
as well as the colour excesses, within the interval given in table 4. 
Physical parameters and abundances were then derived for each one. 
A histogram was built with these values and
the final adopted error for each abundance was the full width of the gaussian
profile fitted to each histogram, taken at half maximum.
These errors are listed in table 6, together with the elemental abundances.

\bigskip
\bigskip
\bigskip
\begin{figure}[h]
\epsfxsize=230pt \epsfbox{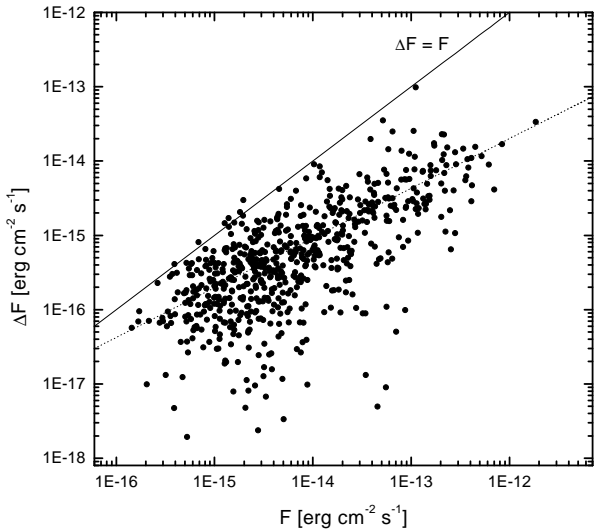}
\caption[]{Errors in line fluxes with respect to their values (in physical units).}
\end{figure}
%

This process was adopted to derive errors in abundances
and temperatures. For densities the value distribution is
not gaussian and errors were derived simply by averaging
those derived from different measures for each object.
 Errors in reddening correspond to the dispersion of those
calculated from each measure.

\begin{table*}
\caption[]{Physical parameters}
\begin{flushleft}
\begin{tabular}{lrrrrrrrrr}
\hline\noalign{\smallskip}
PNe & E(B-V) & $\sigma_{E(B-V)}$ & log(H$\beta$) & n[SII] & $\sigma_{n[SII])}$ 
& T[NII] & $\sigma_{T[NII]}$ & T[OIII] &$\sigma_{T[OIII]}$ \\
\noalign{\smallskip}
\hline\noalign{\smallskip}
K 5-1 & 1.89 & 0.09 & -11.22 & 1.4 & 0.2 & 1.82 & 0.56 & 1.37 & 0.25\\
K 5-3 & 1.27 & 0.13 & -11.82 & 1.3 & 0.4 & 1.07 & 0.23 & 1.38 & 0.15\\
K 5-4 & 1.87 & 0.09 & -10.84 & 7.6 & 1.8 & 1.17 & 0.16 & 0.99 & 0.10\\
K 5-5 & 2.24 & 0.11 & -10.98 & 10.9 & - & 2.94 & 1.43 & 2.08 & 0.46\\
K 5-6 & 1.40 & 0.07 & -11.98 & 1.0 & 0.3 & - & - & 1.82 & 0.25\\
K 5-7 & 1.43 & 0.14 & -12.21 & 0.3 & 0.3 & - & - & 1.75 & 0.30\\
K 5-9 & 2.03 & 0.10 & -11.23 & 3.5 & 1.5 & 1.01 & 0.24 & - & -\\
K 5-11 & 1.92 & 0.38 & -11.52 & 0.7 & 0.2 & 0.81 & 0.10 & 0.94 & 0.20\\
K 5-12 & 1.27 & 0.06 & -11.76 & 5.5 & 1.6 & 1.23 & 0.24 & 1.07 & 0.09\\
K 5-13 & 1.82 & 0.18 & -11.65 & 1.2 & 0.1 & 0.90 & 0.11 & 1.36 & 0.25\\
K 5-14 & 1.42 & 0.14 & -11.67 & 8.7 & - & 1.24 & 0.17 & 1.60 & 0.17\\
K 5-16 & 1.98 & 0.40 & -11.42 & 0.7 & 0.1 & 0.90 & 0.10 & - & -\\
K 5-17 & 1.53 & 0.08 & -11.42 & 11.6 & - & - & - & 1.47 & 0.15\\
K 5-19 & 1.85 & 0.19 & -11.68 & 1.0 & - & 1.49 & 0.41 & 2.48 & 0.57\\
K 5-20 & 1.41 & 0.14 & -11.83 & 0.5 & - & - & - & 1.10 & 0.15\\
SB 01 & 0.71 & 0.07 & -12.99 & 1.4 & 1.3 & - & - & 1.33 & 0.35\\
SB 02 & 0.63 & 0.06 & -12.63 & 0.2 & 0.1 & 1.41 & 0.22 & 1.46 & 0.17\\
SB 03 & 0.52 & 0.05 & -12.67 & 0.3 & - & 1.50 & 0.26 & 3.09 & 0.38\\
SB 04 & 0.43 & 0.04 & -13.26 & 0.1 & 0.1 & - & - & 1.86 & 0.27\\
SB 06 & 0.47 & 0.05 & -13.30 & 0.1 & - & - & - & 2.20 & 0.37\\
SB 12 & 0.11 & 0.11 & -14.58 & - & - & - & - & 2.35 & 0.61\\
SB 15 & 0.42 & 0.21 & -13.07 & - & - & - & - & 2.08 & 0.23\\
SB 17 & 0.58 & 0.06 & -13.25 & 0.1 & - & 1.22 & 0.19 & - & -\\
SB 18 & 0.31 & 0.09 & -13.49 & 0.3 & - & - & - & 1.82 & 0.22\\
SB 19 & 0.49 & 0.05 & -13.18 & 1.8 & - & - & - & 0.90 & 0.11\\
SB 20 & 0.40 & 0.08 & -13.27 & - & - & - & - & 2.58 & 0.29\\
SB 21 & 0.47 & 0.05 & -12.96 & - & - & - & - & 1.70 & 0.20\\
SB 24 & 0.37 & 0.02 & -13.68 & 9.1 & - & - & - & 1.45 & 0.25\\
SB 25 & 1.05 & 0.11 & -12.86 & 0.2 & 0.1 & - & - & 1.30 & 0.29\\
SB 26 & 0.60 & 0.03 & -12.70 & - & - & - & - & 1.62 & 0.16\\
SB 28 & 0.51 & 0.05 & -13.44 & - & - & - & - & 2.90 & 0.64\\
SB 30 & 0.47 & 0.02 & -12.05 & 5.5 & 2.8 & - & - & 1.42 & 0.07\\
SB 31 & 0.70 & 0.04 & -12.40 & 1.0 & 0.5 & - & - & 1.58 & 0.14\\
SB 32 & 0.18 & 0.04 & -13.70 & 0.2 & 0.2 & 1.72 & 0.40 & 1.82 & 0.25\\
SB 33 & 0.50 & 0.05 & -13.56 & 0.4 & 0.1 & 2.06 & 0.59 & 2.18 & 0.38\\
SB 34 & 0.56 & 0.11 & -13.23 & - & - & - & - & 2.39 & 0.44\\
SB 35 & 0.49 & 0.05 & -12.76 & - & - & - & - & 1.18 & 0.10\\
SB 37 & 0.95 & 0.05 & -11.91 & 0.4 & 0.1 & - & - & 1.78 & 0.13\\
SB 38 & 0.04 & 0.04 & -13.65 & - & - & - & - & 1.85 & 0.19\\
SB 42 & 0.48 & 0.02 & -12.35 & 0.1 & 0.1 & 1.31 & 0.28 & 1.24 & 0.11\\
SB 44 & 0.11 & 0.06 & -13.73 & 0.9 & 0.3 & - & - & 1.19 & 0.15\\
SB 50 & 0.43 & 0.09 & -13.12 & 0.3 & 0.3 & - & - & 1.86 & 0.21\\
SB 52 & 0.32 & 0.03 & -12.76 & 1.0 & 1.2 & 1.46 & 0.48 & 1.46 & 0.10\\
SB 53 & 0.50 & 0.05 & -12.63 & 0.7 & 0.1 & 2.53 & 0.74 & 1.74 & 0.13\\
SB 55 & 0.46 & 0.23 & -12.95 & 1.7 & - & - & - & 2.56 & 0.50\\
\noalign{\smallskip}
\hline
\end{tabular}
\end{flushleft}
\end{table*}

We noted from our dereddened fluxes that for some objects the H$\gamma$
flux is considerably different from the recombination value. This effect
is stronger for low latitude objects, indicating a probable relation
with variations in the total-to-selective absorption ratio R$_V$ for some
light paths towards the bulge.  Stasi\'nska et al. (1992) have already mentioned
the possibility of different R$_V$ values for bulge planetary nebulae.
Variations in R$_V$, as pointed out by
Cardelli et al. (1989) can have considerable effects in the blue-UV region.
As the [OIII]$\lambda$4363 line is close to H$\gamma$ and important to 
determine electron temperatures, it was necessary
to check how this effects our derived physical parameters, and
therefore the chemical abundances.
To verify, we scaled the [OIII]$\lambda$4363
line using the ratio between recombination and measured value for
H$\gamma$ as a correction factor, and rederived electron temperatures
and chemical abundances for all objects.
Figure 2 displays a comparison between original and rederived  oxygen abundances. 
As can be seen, variations are within our expected errors.   
Chemical abundances were then rederived using these "scaled" temperatures and
the procedure described below, and the results were found to be consistent
with the original values within the expected errors.

\bigskip
\bigskip
\bigskip
\begin{figure}[h]
\epsfxsize=230pt \epsfbox{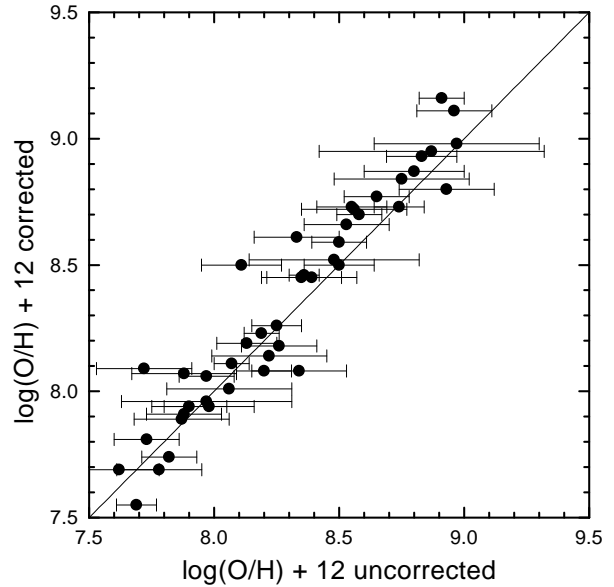}
\caption[]{Comparison between measured and rescaled oxygen abundances
(see section 2.2).}
\end{figure}
%

\section{Chemical abundances}

Ionic abundances were calculated for the ions present in the optical
spectra by solving the statistical equilibrium equations for a three-level
atom model, including radiative and collisional transitions. Elemental
abundances were then derived adopting ionisation correction factors (icf's)
to account for unobserved ions of each element.
Atomic data and icf's used in this work were the same as those adopted by
Costa et al. (1996). Derived ionic abundances are listed in table 5, and the 
resulting elemental abundances are given in table 6.

Oxygen abundance requires ionic abundances of O$^+$ and O$^{++}$. While there
are strong [OIII] lines at $\lambda$4959+5007 \AA, for O$^+$  we have in
principle two possibilities: the blue pair [OII]$\lambda$3626+29 \AA, and
the red one [OII]$\lambda$7320+20 \AA. In this work we adopted the red pair
due to the greater sensitivity to reddening of the blue-UV region,
as discussed in section 2.2, and due to the greater efficiency of the 
instrumental set in the red region. For many objects we were not able to
get flux for the $\lambda$3626+29 pair due to low S/N. In view of this fact 
we chose to keep the same line for all objects.
Only for a few objects, as can be seen in tables 2 and 3, do we not have the red
pair and were forced to adopt the blue one to derive O$^+$  abundance.

To verify an eventual dependence of the line adopted on the O$^+$ abundance,
for those objects having both blue and red lines measured we calculated
the ionic abundance from both and compare them. The result is shown
in figure 3. As can be seen, both values are perfectly compatible, and
we expect no differences in the final oxygen abundance when using  blue
or red [OII] lines.

\begin{figure}[h]
\vspace{1cm}
\epsfxsize=230pt \epsfbox{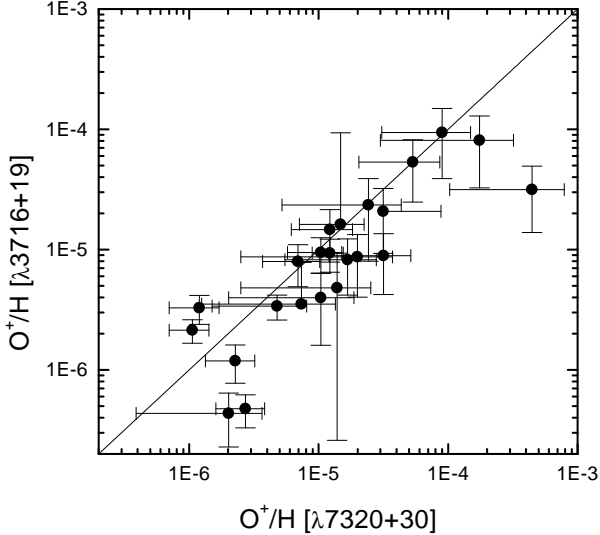}
\caption[]{O$^+$ ionic abundance derived from blue and red [OII] pairs
(see text). Both values are clearly compatible, the diagonal x=y line was drawn 
over the points and does not represent any adjustment. }
\end{figure}
%

Derivation of helium abundances requires special caution because collisional
effects are present. In the literature there are correction factors to
be applied to HeI abundances, however they may be overestimated up to
a factor of two (Peimbert \& Torres-Peimbert (1987). Here we adopted the
recombination theory for HeI and HeII, using recombination coefficients 
from P\'equignot (1991). HeI abundance was also corrected for collisional effects
using the correction terms from Kingdon \& Ferland (1995).

Helium abundance is given as
He/H and other elemental abundances are given as $\varepsilon$(X)=log(X/H)+12.
Errors in the abundances refer always to X/H.

\begin{table*}
\caption[]{Ionic abundances}
\begin{flushleft}
\begin{tabular}{lrrrrrrrr}
\hline\noalign{\smallskip}
Name & HeI & HeII & NII & SII & SIII & OII & OIII & ArIII\\
\noalign{\smallskip}
\hline\noalign{\smallskip}
K 5-1 & 0.108 & - & 9.29E-07 & 2.69E-08 & 4.57E-07 & 2.47E-06 & 7.29E-05 & 5.76E-07\\
K 5-3 & 0.082 & 0.042 & 4.19E-06 & 1.82E-07 & 9.44E-07 & 3.16E-05 & 1.17E-04 & 4.06E-07\\
K 5-4 & 0.109 & - & 1.02E-05 & 3.31E-07 & 3.97E-06 & 5.34E-05 & 3.91E-04 & 1.12E-06\\
K 5-5 & 0.03 & - & 5.63E-07 & 4.62E-08 & 2.93E-07 & 2.06E-06 & 5.08E-05 & 2.02E-07\\
K 5-6 & 0.023 & 0.09 & 4.06E-07 & 4.16E-08 & 5.59E-07 & 2.22E-06 & 7.07E-05 & 2.09E-07\\
K 5-7 & 0.066 & 0.094 & 1.25E-06 & 7.38E-08 & - & 5.22E-06 & 8.44E-05 & 6.10E-07\\
K 5-9 & 0.142 & 0.015 & 7.73E-06 & 2.50E-07 & 1.29E-06 & 4.17E-05 & 2.32E-04 & 1.14E-06\\
K 5-11 & 0.174 & 0.017 & 1.13E-04 & 2.36E-06 & 3.19E-06 & 4.57E-04 & 2.22E-04 & 1.64E-06\\
K 5-12 & 0.046 & 0.06 & 2.91E-06 & 2.05E-07 & 3.11E-06 & 1.25E-05 & 3.45E-04 & 5.61E-06\\
K 5-13 & 0.082 & 0.058 & 8.68E-05 & 2.31E-06 & 4.20E-06 & 1.75E-04 & 1.95E-04 & 1.53E-06\\
K 5-14 & 0.094 & 0.022 & 2.50E-05 & 8.56E-07 & 1.68E-06 & 8.88E-05 & 1.71E-04 & 9.26E-07\\
K 5-16 & 0.129 & 0.012 & 1.28E-04 & 4.13E-06 & 1.78E-05 & 4.12E-04 & 4.47E-04 & 2.32E-06\\
K 5-17 & 0.076 & - & 1.77E-06 & 1.73E-07 & 6.41E-07 & 6.97E-06 & 1.52E-04 & 3.66E-07\\
K 5-19 & 0.071 & 0.082 & 3.34E-06 & 1.73E-07 & - & 7.57E-06 & 5.29E-05 & 2.30E-07\\
K 5-20 & 0.128 & - & 7.30E-07 & 3.08E-08 & 7.78E-07 & 3.17E-05 & 2.14E-04 & 4.74E-07\\
SB 01 & 0.046 & 0.032 & 1.79E-06 & 1.68E-07 & - & 1.48E-05 & 8.26E-05 & 1.70E-07\\
SB 02 & 0.141 & 0.027 & 6.96E-06 & 2.48E-07 & - & 1.99E-05 & 5.90E-05 & 3.62E-07\\
SB 03 & 0.105 & 0.054 & 1.59E-06 & 6.55E-08 & - & 1.24E-06 & 3.09E-05 & 1.78E-07\\
SB 04 & 0.095 & 0.019 & 6.34E-06 & 1.75E-07 & - & 6.12E-06 & 3.86E-05 & 1.85E-07\\
SB 06 & 0.084 & 0.046 & 2.18E-05 & 1.27E-07 & - & 1.22E-05 & 2.65E-05 & 3.72E-07\\
SB 12 & 0.049 & 0.066 & 2.93E-07 & - & - & - & 4.08E-05 & 2.52E-07\\
SB 15 & 0.053 & 0.046 & 7.64E-08 & 6.16E-08 & - & 4.78E-06 & 7.79E-05 & 4.21E-08\\
SB 17 & 0.142 & - & 3.06E-05 & 1.45E-06 & - & 8.98E-05 & 2.58E-06 & 5.47E-08\\
SB 18 & 0.046 & 0.057 & 1.98E-07 & 2.25E-08 & - & 6.93E-06 & 7.53E-05 & 6.85E-08\\
SB 19 & 0.156 & - & 2.09E-06 & 2.48E-07 & 5.05E-06 & 3.16E-05 & 3.34E-04 & 1.85E-06\\
SB 20 & 0.116 & 0.028 & 5.90E-08 & 2.64E-08 & - & 1.37E-06 & 6.02E-05 & 4.48E-08\\
SB 21 & 0.021 & 0.102 & 1.02E-07 & - & - & 3.11E-06 & 5.04E-05 & 2.01E-07\\
SB 24 & 0.032 & 0.077 & 5.68E-07 & 1.56E-07 & - & 1.67E-05 & 4.79E-05 & 1.62E-07\\
SB 25 & 0.138 & - & 3.24E-05 & 9.02E-07 & 8.16E-07 & 2.94E-05 & 8.54E-05 & 8.10E-07\\
SB 26 & 0.016 & 0.082 & 2.19E-07 & - & 1.45E-07 & 1.20E-06 & 8.96E-05 & 1.50E-07\\
SB 28 & 0.004 & 0.11 & 3.86E-08 & - & - & - & 1.84E-05 & 5.70E-08\\
SB 30 & 0.061 & 0.04 & 1.92E-06 & 6.88E-08 & 3.46E-07 & 1.22E-05 & 1.27E-04 & 1.87E-07\\
SB 31 & 0.009 & 0.101 & 3.33E-07 & 2.17E-08 & 8.03E-07 & 2.27E-06 & 6.84E-05 & 1.59E-07\\
SB 32 & 0.12 & 0.022 & 6.71E-06 & 4.40E-07 & 2.66E-06 & 2.43E-05 & 4.32E-05 & 2.05E-07\\
SB 33 & 0.079 & 0.016 & 7.51E-06 & 3.55E-07 & - & 1.04E-05 & 2.43E-05 & 2.85E-07\\
SB 34 & 0.01 & 0.129 & 4.09E-07 & - & - & - & 2.37E-05 & 7.48E-08\\
SB 35 & 0.107 & - & 6.09E-07 & 4.24E-08 & - & 1.04E-05 & 1.67E-04 & 3.87E-07\\
SB 37 & - & 0.146 & 7.93E-07 & 4.91E-08 & 1.13E-06 & 1.06E-06 & 9.73E-05 & -\\
SB 38 & 0.077 & 0.058 & 1.24E-07 & 3.83E-08 & - & 4.76E-07 & 6.68E-05 & 1.40E-07\\
SB 42 & 0.051 & 0.004 & 1.08E-06 & 4.22E-08 & - & 1.39E-05 & 4.70E-05 & 1.52E-07\\
SB 44 & 0.029 & 0.107 & 3.25E-06 & 1.39E-07 & - & 5.30E-05 & 1.27E-04 & 6.99E-07\\
SB 50 & 0.056 & 0.086 & 7.41E-07 & 8.29E-08 & - & 3.56E-06 & 5.02E-05 & 2.73E-07\\
SB 52 & 0.033 & 0.103 & 3.93E-07 & 5.52E-08 & 6.56E-07 & 7.37E-06 & 8.44E-05 & 4.09E-07\\
SB 53 & 0.015 & 0.119 & 5.56E-07 & 3.81E-08 & 1.45E-06 & 1.10E-06 & 7.61E-05 & 4.75E-07\\
SB 55 & 0.035 & 0.118 & 1.08E-07 & 3.64E-09 & - & 2.02E-06 & 1.50E-05 & 8.31E-08\\
\noalign{\smallskip}
\hline\noalign{\smallskip}
\noalign{\smallskip}
\end{tabular}
\end{flushleft}
\end{table*}

\begin{table*}
\caption[]{Elemental abundances}
\begin{flushleft}
\begin{tabular}{lrrrrrrrrrrr}
\hline\noalign{\smallskip}
Planetary & He & $\sigma_{He}$ & $\varepsilon$(N) & $\sigma_{N}$ & $\varepsilon$(S) & 
$\sigma_{S}$ & $\varepsilon$(O) & $\sigma_{O}$ & $\varepsilon$(Ar) & $\sigma_{Ar}$ & Notes\\
\noalign{\smallskip}
\hline\noalign{\smallskip}
K 5-1 & 0.108 & 0.031 & 7.45 & 0.30 & 6.91 & 0.68 & 7.88 & 0.21 & 5.9 & 0.16\\
K 5-3 & 0.123 & 0.019 & 7.48 & 0.20 & 6.45 & 0.27 & 8.35 & 0.16 & 6.02 & 0.17\\
K 5-4 & 0.109 & 0.014 & 7.93 & 0.17 & 7.24 & 0.28 & 8.65 & 0.13 & 6.23 & 0.14\\
K 5-5 & 0.030 & 0.013 & 7.16 & 0.24 & 6.67 & 0.48 & 7.72 & 0.19 & 5.45 & 0.17 & 1\\
K 5-6 & 0.114 & 0.017 & 7.81 & 0.21 & 7.05 & 0.18 & 8.55 & 0.14 & 6.15 & 0.15\\
K 5-7 & 0.160 & 0.031 & 7.72 & 0.14 & 5.8 & 0.17 & 8.33 & 0.17 & 6.32 & 0.18\\
K 5-9 & 0.157 & 0.030 & 7.75 & 0.15 & 6.7 & 0.31 & 8.48 & 0.34 & 6.3 & 0.26\\
K 5-11 & 0.191 & 0.056 & 8.26 & 0.22 & 6.92 & 0.38 & 8.87 & 0.45 & 6.87 & 0.67\\
K 5-12 & 0.106 & 0.013 & 8.28 & 0.28 & 7.71 & 0.51 & 8.91 & 0.09 & 7.25 & 0.10\\
K 5-13 & 0.140 & 0.029 & 8.5 & 0.11 & 7.03 & 0.20 & 8.8 & 0.20 & 6.82 & 0.27\\
K 5-14 & 0.116 & 0.019 & 7.95 & 0.11 & 6.68 & 0.15 & 8.5 & 0.14 & 6.37 & 0.22\\
K 5-16 & 0.141 & 0.042 & 8.47 & 0.09 & 7.56 & 0.31 & 8.97 & 0.33 & 6.82 & 0.64\\
K 5-17 & 0.076 & 0.015 & 7.61 & 0.17 & 6.99 & 0.25 & 8.2 & 0.11 & 5.71 & 0.12\\
K 5-19 & 0.153 & 0.031 & 7.76 & 0.25 & 5.82 & 0.20 & 8.11 & 0.16 & 5.88 & 0.22\\
K 5-20 & 0.128 & 0.027 & 6.75 & 0.13 & 6.48 & 0.21 & 8.39 & 0.18 & 5.86 & 0.19\\
SB 01 & 0.077 & 0.016 & 7.3 & 0.16 & 5.74 & 0.16 & 8.22 & 0.23 & 5.66 & 0.19 & 1\\
SB 02 & 0.168 & 0.024 & 7.52 & 0.13 & 5.74 & 0.09 & 7.97 & 0.11 & 5.89 & 0.12\\
SB 03 & 0.159 & 0.020 & 7.8 & 0.11 & 5.96 & 0.1 & 7.69 & 0.08 & 5.58 & 0.09 & 2,3\\
SB 04 & 0.114 & 0.028 & 7.75 & 0.12 & 5.79 & 0.11 & 7.73 & 0.13 & 5.54 & 0.13\\
SB 06 & 0.130 & 0.029 & 8.03 & 0.09 & 5.4 & 0.13 & 7.78 & 0.17 & 6.05 & 0.15\\
SB 12 & 0.115 & 0.034 & - & - & - & - & 7.98 & 0.18 & 5.9 & 0.25\\
SB 15 & 0.099 & 0.017 & 6.39 & 0.17 & 5.72 & 0.13 & 8.19 & 0.07 & 5.05 & 0.19\\
SB 17 & 0.142 & 0.033 & 7.5 & 0.15 & 6.32 & 0.16 & 7.97 & 0.34 & 6.42 & 0.35\\
SB 18 & 0.103 & 0.023 & 6.72 & 0.26 & 5.11 & 0.23 & 8.26 & 0.15 & 5.35 & 0.21\\
SB 19 & 0.156 & 0.031 & 7.38 & 0.19 & 7.47 & 0.28 & 8.56 & 0.21 & 6.43 & 0.16 & 3\\
SB 20 & 0.144 & 0.029 & 6.52 & 0.17 & 5.85 & 0.18 & 7.88 & 0.15 & 4.88 & 0.19 & 3\\
SB 21 & 0.123 & 0.017 & 7.02 & 0.16 & - & - & 8.5 & 0.11 & 6.23 & 0.13\\
SB 24 & 0.109 & 0.018 & 6.87 & 0.30 & 5.53 & 0.37 & 8.34 & 0.19 & 6 & 0.21\\
SB 25 & 0.138 & 0.038 & 8.1 & 0.15 & 6.58 & 0.22 & 8.06 & 0.25 & 6.16 & 0.21 & 3\\
SB 26 & 0.098 & 0.013 & 8 & 0.17 & 6.88 & 0.26 & 8.74 & 0.10 & 6.09 & 0.12\\
SB 28 & 0.114 & 0.021 & - & - & - & - & 8.75 & 0.27 & 6.37 & 0.28\\
SB 30 & 0.100 & 0.010 & 7.56 & 0.12 & 6.35 & 0.17 & 8.36 & 0.06 & 5.66 & 0.06\\
SB 31 & 0.110 & 0.011 & 8.13 & 0.20 & 7.15 & 0.19 & 8.96 & 0.15 & 6.45 & 0.14\\
SB 32 & 0.142 & 0.032 & 7.34 & 0.14 & 6.76 & 0.20 & 7.9 & 0.15 & 5.71 & 0.16\\
SB 33 & 0.095 & 0.029 & 7.48 & 0.13 & 5.85 & 0.12 & 7.62 & 0.16 & 5.82 & 0.16\\
SB 34 & 0.139 & 0.023 & - & - & - & - & 8.53 & 0.17 & 6.16 & 0.19\\
SB 35 & 0.107 & 0.016 & 7.02 & 0.12 & 5.55 & 0.15 & 8.25 & 0.10 & 5.74 & 0.11\\
SB 37 & 0.146 & 0.015 & - & - & 7.91 & 0.17 & - & - & - & -\\
SB 38 & 0.135 & 0.017 & 7.49 & 0.16 & 6.65 & 0.19 & 8.07 & 0.07 & 5.52 & 0.12 & 3\\
SB 42 & 0.055 & 0.010 & 6.71 & 0.19 & 5 & 0.10 & 7.82 & 0.11 & 5.45 & 0.11\\
SB 44 & 0.136 & 0.017 & 7.72 & 0.18 & 5.45 & 0.17 & 8.93 & 0.19 & 6.8 & 0.18\\
SB 50 & 0.141 & 0.021 & 7.45 & 0.15 & 5.79 & 0.17 & 8.13 & 0.12 & 6 & 0.12 & 3\\
SB 52 & 0.137 & 0.011 & 7.31 & 0.34 & 6.63 & 0.59 & 8.58 & 0.09 & 6.39 & 0.10\\
SB 53 & 0.134 & 0.012 & 8.53 & 0.2 & 7.85 & 0.4 & 8.83 & 0.14 & 6.75 & 0.12 & 3\\
SB 55 & 0.153 & 0.026 & 6.6 & 0.32 & 4.17 & 0.40 & 7.87 & 0.19 & 5.74 & 0.24\\
\noalign{\smallskip}
\hline\noalign{\smallskip}
\noalign{\smallskip}
\end{tabular}
\end{flushleft}
\begin{list}{}{}
\item[1] Spectra with large errors
\item[2] An average temperature was used
\item[3] Lines [OII]$\lambda$3726+29 were used to derive OII abundance
\end{list}
\end{table*}

\section {Discussions and Conclusion}

Different samples were used for this work. Most of the objects listed
by Kohoutek (1994) are 
in the northern galactic hemisphere while those from Beaulieu
et al. (1999) are mostly in the southern galactic hemisphere.
Also, we selected additional data from the literature for our analysis. 
These additional abundances are not homogeneous with respect to observation
and reduction procedures, however they are important for
the sampling of the range of elemental abundances. There are
few works on bulge PNe abundances, as this kind of result requires a
good S/N ratio towards a region of high extinction. Three
works were then selected:
Cuisinier et al. (2000), Webster (1988) and Aller \& Keyes (1987). 
Webster's abundances were re-derived using the same icf's as we
used for ours.
For each of these samples we selected only objects with good spectra.

Table 6 lists our final abundances for each object. Mean values and their
distribution were also derived for the sample, where 
distribution is represented
by a 2$\sigma$ deviation from the mean value. Final values
are: He: 0.126 $\pm$ 0.054; N: 7.64 $\pm$ 1.10; O: 8.22 $\pm$ 0.85; 
S: 6.48 $\pm$ 1.87; Ar: 5.95 $\pm$ 1.01.

These values represent averages and FWHM of gaussians adjusted
to each histogram of abundance distribution. 
It should be noted that nitrogen elemental abundances depend on OII 
ionic abundances used as icf, so uncertainties in this last parameter 
will increase the dispersion found in N. As nitrogen is also a product of
stellar nucleosynthesis dredged-up during the stellar evolution
of the progenitors, their mass spectrum will also contribute to
the derived dispersion in N abundance.

Sulphur abundances have to be taken cautiously because they were
derived from relatively weak [SII]$\lambda$6716+31 and [SIII]
$\lambda$6312 lines. Besides, sulphur abundances in particular
are strongly dependent on electron temperatures.
\begin{figure}[h]
\vspace{1cm}
\epsfxsize=230pt \epsfbox{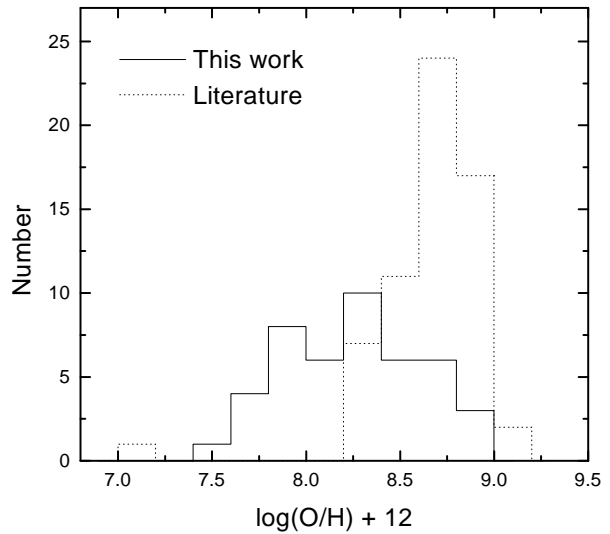}
\caption[]{Distribution of oxygen abundances for our sample (full line)
 and for the literature (dashed line).}
\end{figure}
%

\begin{figure}[h]
\vspace{1.5cm}
\epsfxsize=230pt \epsfbox{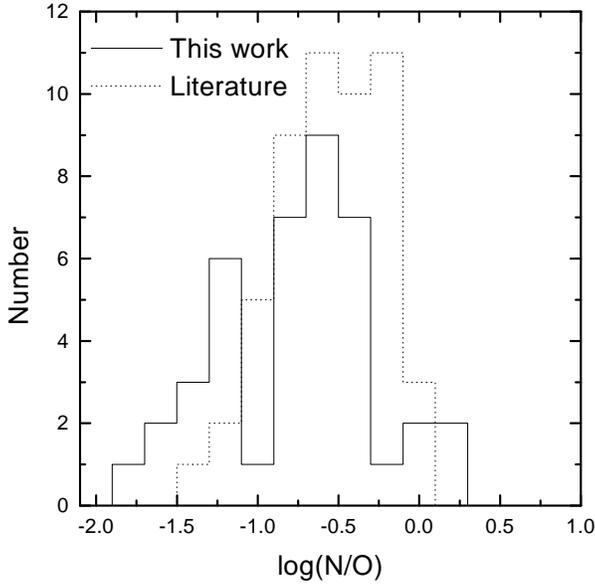}
\caption[]{Distribution of log (N/O) for our sample (full line)
 and for the literature (dashed line).}
\end{figure}
%
\begin{figure}[h]
\vspace{1cm}
\epsfxsize=230pt \epsfbox{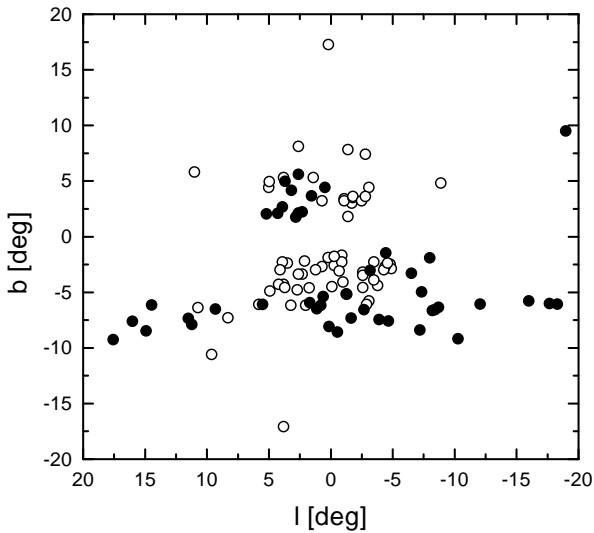}
\caption[]{Space distribution for our sample (filled circles) and
for the literature (open circles).}
\end{figure}
%
\begin{figure}[h]
\vspace{1.5cm} 
\epsfxsize=230pt \epsfbox{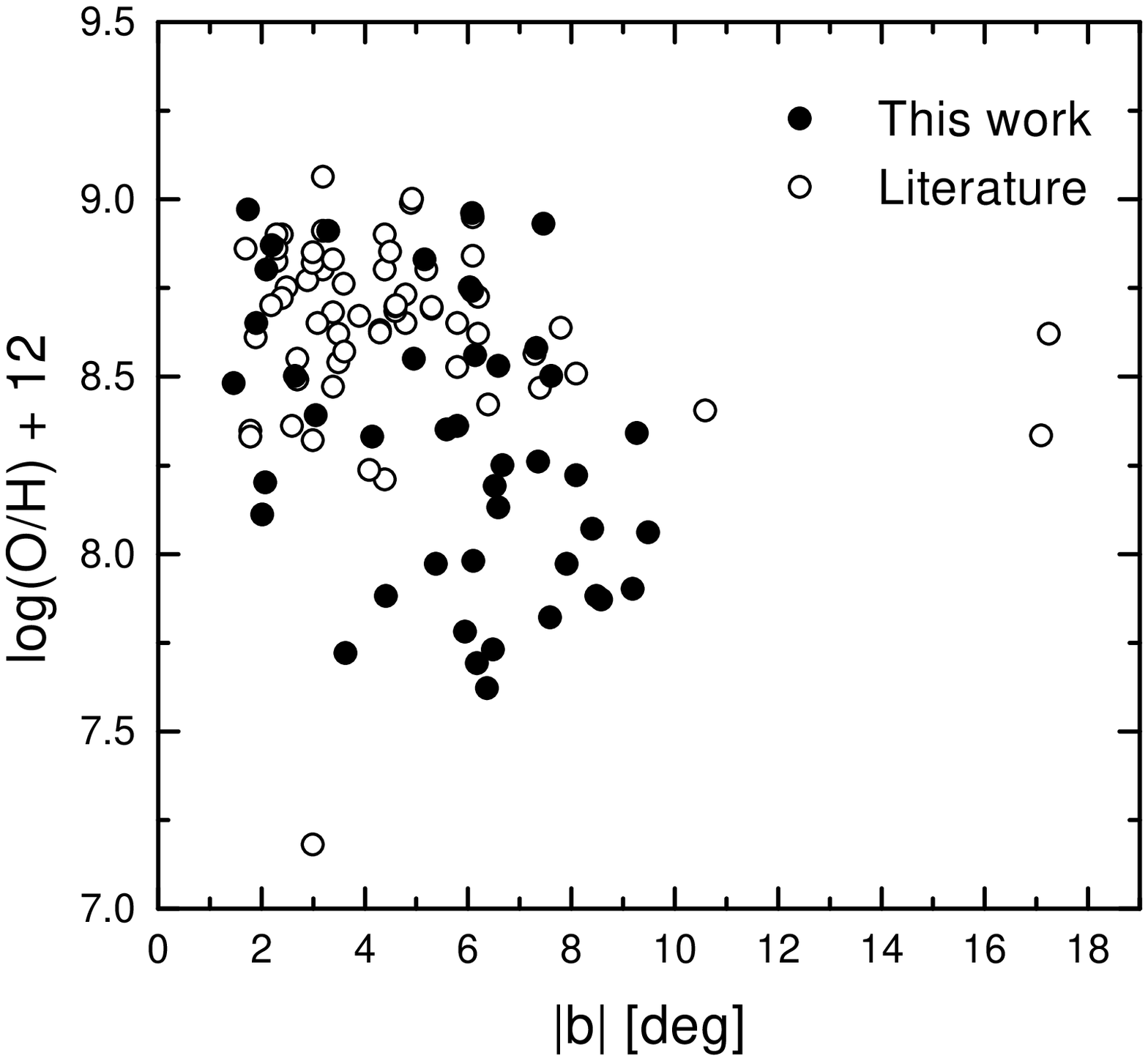}
\caption[]{Oxygen abundances vs. absolute value of galactic latitude
for our sample (filled circles) and for the literature (open circles).}
\end{figure}
%

Figures 4 and 5 
display the abundance distribution for our sample compared
to that from other works, respectively for oxygen and log (N/O). 
At a first glance, distributions seem to be
different, but this is due to different coverages in the sky. 
Our sample spans over 20 deg. in galactic longitude and 11 deg.
in latitude, in a nearly homogeneous distribution, while data from the 
literature are strongly concentrated nearer the position of
the galactic center. These different distributions can be seen
in fig. 6, where both samples are mapped in galactic co-ordinates. 

N/O distribution reflects the mass distribution of the 
progenitors, with more massive objects having higher N abundances.
A few objects with log(N/O)$<-$1 could be originated from old, small-mass
progenitors. The distribution peaks for typical type-II PNe abundances. 

Concerning abundances, fig. 7 displays the oxygen abundance distribution in latitude
for our sample, compared with results from the literature taken from the
three works mentioned before. Examining the results, we see in our sample
a concentration of objects around $\varepsilon$(O)=8.0 at latitudes
greater than 5 deg., which deviates from the average distribution.
These objects belong to a region where no PNe abundances have been 
determined before. 
They can be thick disk objects seen towards the bulge, or effectively
bulge objects at higher altitudes above the galactic plane. If their
membership is confirmed, their lower abundances suggest the
existence of a vertical abundance gradient within the bulge.

Figure 8 shows log(N/O) versus helium abundance for our sample and
for the same literature objects used above. As nitrogen and helium
are nucleosynthesis products and related
to the mass spectrum of the progenitors, they are expected to be correlated.
In the figure there is an upper group of objects that, in spite of
the expected dispersion due to the mass distribution of the progenitors,
shows a correlation. 
It can also be seen that there
is another small group with low N/O and He/H varying from 0.10 to 0.16. The
same pattern appears in the results of Cuisinier et al. (1996) and can
be related to objects at high $z$ above the galactic plane. However, as
distances for our these object are highly uncertain, this is still an
open question.


\begin{figure}[h]
\vspace{1cm} 
\epsfxsize=230pt \epsfbox{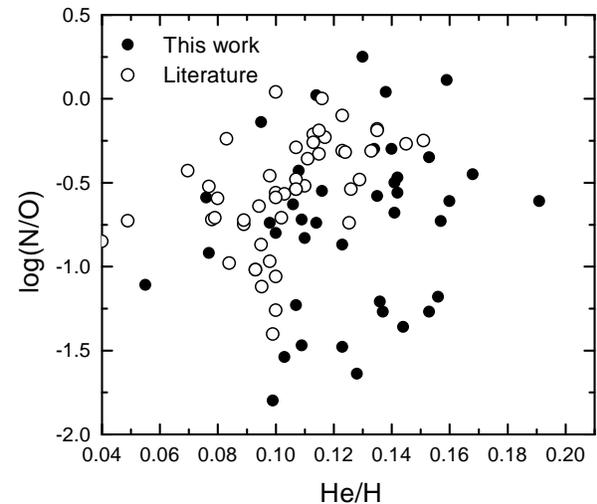}
\caption[]{log(N/O) vs. helium abundance 
for our sample (filled circles) and for the literature (open circles).}
\end{figure}
%

One has to take into account that bulge membership is an 
additional problem hard to solve for PNe. Objects from 
Beaulieu et al. (1999) are distributed over a wide range
of galactic latitudes and their elemental abundances are
lower than those from other works. This problem will be
tackled in another work, now in preparation.

\begin{acknowledgements}
A.V.E. acknowledges CAPES for his graduate fellowship. Observations
at ESO/Chile were possible through FAPESP grant 98/10138-8. Comments
and suggestions of the referee helped to improve the final quality
of this work.
\end{acknowledgements}


\begin{thebibliography}{}

\bibitem[]{Acker} Acker A., Ochsenbein F., Stenholm B.,
Tylenda R., Marcout J., Schohn C. 1992,
Strasbourg-ESO Catalogue of Galactic Planetary Nebulae
\bibitem[]{}Aller,L.H., Keyes,C.D. 1987, ApJ Suppl. 65, 405
\bibitem[]{Beaulieu} Beaulieu S. F., Dopita M. A., Freeman K. C. 1999,
ApJ. 515, 610
\bibitem[]{}Cardelli,J.A., Clayton,G.C., Mathis,J.S. 1989, ApJ 345, 245 
\bibitem[]{Costa} Costa R. D. D., Chiappini C., Maciel W. J.,
de Freitas Pacheco J. A. 1996, A\&A Suppl. 116, 249
\bibitem[]{}Costa,R.D.D., Idiart,T.P., de Freitas Pacheco, J.A. 2000, 
A\&A Supp. 145, 467
\bibitem[]{}Cuisinier,F., Acker,A., K\"oppen,J. 1996, A\&A 307, 215
\bibitem[]{}Cuisinier,F., Maciel,W.J., K\"oppen,J., Acker,A., Stenholm,B. 2000,
A\&A 353, 543
\bibitem[1992]{Freitas} Freitas Pacheco J. A., Maciel W. J., Costa R.D.D.
1992, A\&A 261, 579
\bibitem[]{}Frogel,J.A., Terndrup,D.M., Blanco,V.M., Whitford,A.E. 1990, ApJ
353, 494
\bibitem[]{}Frogel,J.A., Tiede,G.P., Kuchinski,L.E. 1999, AJ 117, 2296
\bibitem[]{}Ibata, R.A., Gilmore, G.F. 1995, MNRAS 275, 605
\bibitem[]{}Idiart,T.P., de Freitas Pacheco,J.A., Costa, R.D.D. 1996, AJ 111, 1169
\bibitem[]{}Kingdon,J., Ferland,G.J. 1995, ApJ 442, 714
\bibitem[1994]{Kohoutek} Kohoutek L. 1994, Ast. Nach. 315, 235
\bibitem[]{}McWilliam,A. 1997, Ann.Rev.A\&A 35, 503
\bibitem[]{}McWilliam,A., Rich,R.M. 1994, ApJ.Suppl. 91, 749
\bibitem[]{}Minitti,D., Olszewski,E.W., Liebert,J., White,S.D.M., Hill,J.M.,
Irwin,M.J. 1995, MNRAS 277, 1293
\bibitem[1989]{Osterbrock} Osterbrock D.E. 1989,
in: Astrophysics of Gaseous Nebulae and Active Galactic Nuclei, University
Science Books
\bibitem[]{} Peimbert,M., Torres-Peimbert,S. 1987, Rev.Mex.A\&A 14, 540
\bibitem[]{} P\'equignot,D., Petitjean,P., Boisson,C. 1991, A\&A 251, 680 
\bibitem[]{} Rich, R.M. 1988, AJ 95, 82
\bibitem[]{} Ratag,M.A., Pottasch,S.R., Dennefeld,M., Menzies, J.W. 1992,
A\&A 255, 255
\bibitem[]{} Ratag,M.A., Pottasch,S.R., Dennefeld,M., Menzies, J.W. 1997,
A\&A Suppl. 126, 297
\bibitem[]{} Stasi\'nska,G., Tylenda,R., Acker,A., Stenholm,B. 1992, A\&A 266, 486 
\bibitem[]{} Tiede,G.M., Frogel,J.A., Terndrup,D.M. 1995, AJ 110, 2788
\bibitem[]{} van den Bergh,S. 1996, PASP 108, 1091
\bibitem[]{} Webster,L. 1988, MNRAS 230, 377
\end{thebibliography}
\end{document}